\documentclass[aps,prb,twocolumn,floatfix]{revtex4}
\usepackage[final]{graphicx}
\usepackage{bm}
\usepackage{float}
\usepackage{afterpage}

\begin{document}

\title{Fractional topological phase in one-dimensional flatbands with
nontrivial topology}
\author{Huaiming Guo$^{1}$, Shun-Qing Shen$^2$ and Shiping Feng$^3$}
\affiliation{$^1$Department of Physics, Beihang University, Beijing, 100191, China}
\affiliation{$^2$Department of Physics, The University of Hong Kong, Pokfulam Road, Hong
Kong}
\affiliation{$^3$Department of Physics, Beijing Normal University, Beijing, 100875, China}

\begin{abstract}
We show the existence of the fractional topological phase (FTP) in a
one-dimensional interacting fermion model using exact diagonalization, in
which the non-interacting part has flatbands with nontrivial topology. In
the presence of the nearest-neighbouring interaction $V_{1}$, the FTP at
filling factor $\nu =1/3$ appears. It is characterized by the three-fold
degeneracy and the quantized total Berry phase of the ground-states. The FTP
is destroyed by a next-nearest-neighbouring interaction $V_{2}$ and the
phase diagrams in the $(V_{1},V_{2})$ plane is determined. We also present a
physical picture of the phase and discuss its existence in the nearly
flatband. Within the picture, we argue that the FTP at other filling factors
can be generated by introducing proper interactions. The present study
contributes to a systematic understanding of the FTPs and can be realized in
cold-atom experiments.
\end{abstract}

\pacs{
  73.43.Cd,   05.30.Fk,   71.10.Fd,             }
\maketitle

\section{Introduction}
The discovery of integer (IQHE) and fractional
(FQHE) quantum Hall effects opened a window to explore the mystery in
condensed matter physics \cite{iqhe,fqhe,fqhe1}. They are new topological
states of quantum matter, which go beyond the Landau's theory of spontaneous
symmetry breaking \cite{landau}. The studies of these effects enrich our
understanding of quantum phases and quantum phase transitions. In his
seminal paper \cite{haldane}, Haldane showed that IQHE can be realized on a
lattice model without a net magnetic field. Most recent generalization of
the Haldane model to electrons with spin 1/2 gives birth to the
time-reversal invariant $Z_{2}$ topological insulator (TI), which becomes
the current research focus in condensed matter physics due to their many
exotic properties \cite{kane1,kane2,rmp1,rmp2,rmp3}.

Inspired by this, similar ideas arise for the situation of FQHE and many
studies are devoted to realize the fractional topological phase (FTP) on
lattice models in the absence of external magnetic fields. Up to now great
development has been achieved. Models that exhibit nearly flatband with a
nonzero Chern number are proposed in different systems and numerical
calculations confirm its existence when interactions are included \cite%
{flat1,flat2,flat3,flat4,flat5,flat6,flat7,flat8,inter1,inter2}. To gain
some insight of FTP, in this paper we focus on the one-dimensional (1D)
case. Using exact diagonalization, we calculate the low-energy spectrum and
Berry phase of the lowest energy states of a 1D interacting topological flat
band model, and identify the FTP at the filling factor, $i.e.$, the average
number of electron per site, $\nu =1/3$. We also present a physical picture
of the phase and discuss its possible existence of fractional charges in the
nearly flatband. The FTPs at other filling factors can be generated by
introducing proper interactions.

\begin{figure}[tbp]
\includegraphics[width=8cm]{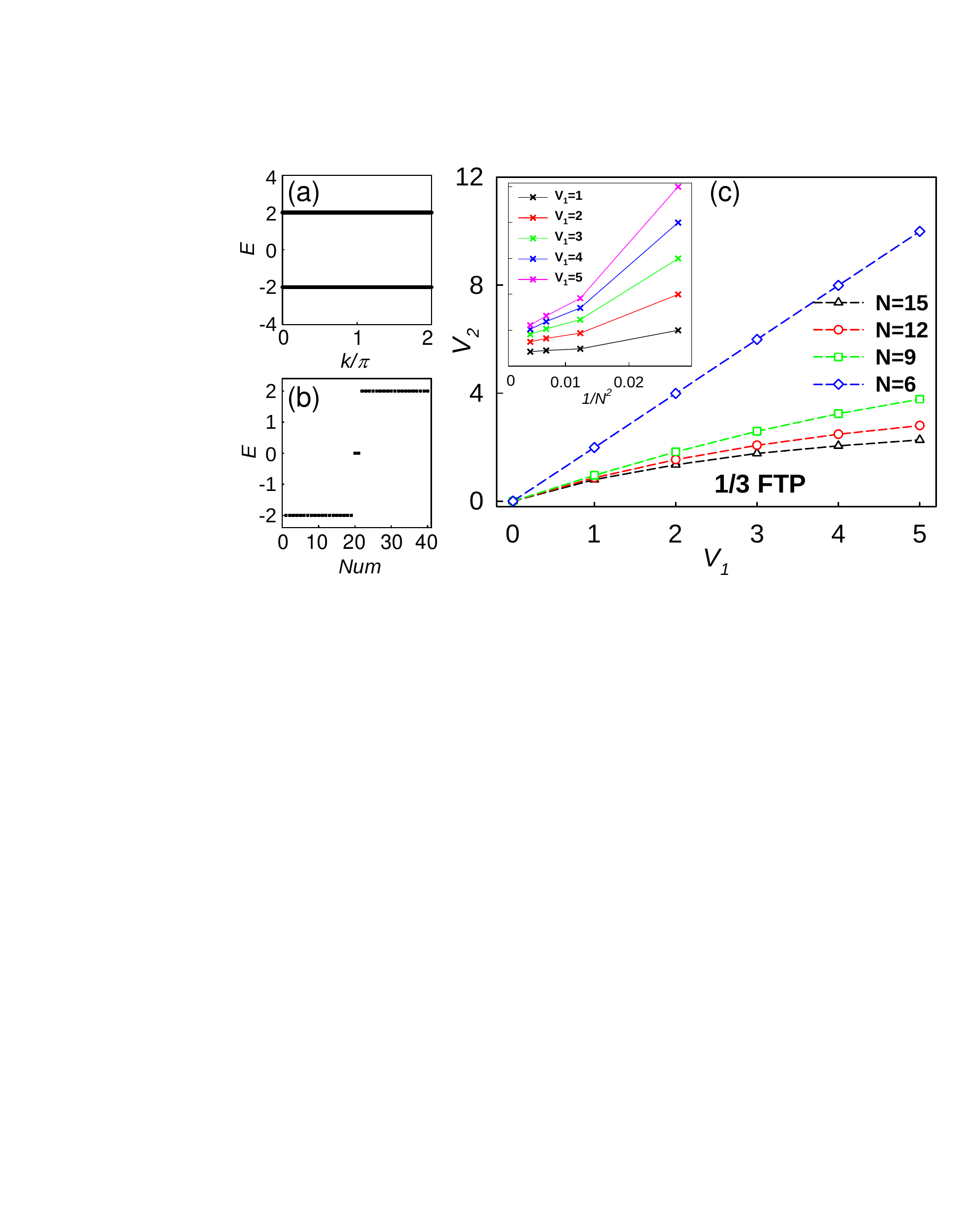}
\caption{(Color online) (a)The tight-binding band structure when the
parameters satisfy $-M/2=B=A$, where the bands are flat. (b) Edge modes in
the flatband case on a chain of length $N=20$ with open boundary condition.
(c) The phase diagram in the $(V_1,V_2)$ plane at $\protect\nu=1/3$ for
different sizes. The inset shows the size dependence of the critical value $%
V_{2c}$ at different $V_1$. Here $A=B=1$ and $M=-2$ (in the following
calculations if not mentioned, we use these parameters).}
\label{fig1}
\end{figure}

\section{1D topological flatband model}
Consider the 1D non-interacting
tight-binding model \cite{model},
\begin{eqnarray}
H_{0} &=&\sum_{i}(M+2B)\Psi _{i}^{\dagger }\sigma _{z}\Psi _{i}-\sum_{i,\hat{%
x}}B\Psi _{i}^{\dagger }\sigma _{z}\Psi _{i+\hat{x}}  \nonumber  \label{eq1}
\\
&-&\sum_{i,\hat{x}}sgn(\hat{x})iA\Psi _{i}^{\dagger }\sigma _{x}\Psi _{i+%
\hat{x}}
\end{eqnarray}%
where $\sigma _{x}$ and $\sigma _{z}$ are Pauli matrices, and the spinor $%
\Psi _{i}=(c_{i},d_{i})^{T}$ with $c_{i}$ ($d_{i}$) electron annihilating
operator at the site $\mathbf{r}_{i}$, which can be obtained by mapping the
Dirac equation into a lattice \cite{shen}. In the momentum space Eq.(\ref%
{eq1}) becomes $H_{0}=\sum_{k}\Psi _{k}^{\dagger }\mathcal{H}(k)\Psi _{k}$
with $\Psi _{k}=(c_{k},d_{k})^{T}$ the Fourier partner of $\Psi _{i}$ and
\[
\mathcal{H}(k)=[M+2B-2B\cos (k)]\sigma _{z}+2A\sin (k)\sigma _{x}.
\]%
The spectrum of $\mathcal{H}(k)$ consists of two bands,
\[
E_{k}^{(1,2)}=\pm \sqrt{\lbrack M+2B-2B\cos (k)]^{2}+[2A\sin (k)]^{2}}.
\]%
Usually the two bands are dispersive, but when the parameters satisfy $%
-M/2=B=A$, the bands are flat, when $E_{k}=\pm |2A|$. Also the resulting
flatband has nontrivial Berry phase, $\pi $ ${mod}(2\pi ),$ which is
manifested by the existence of the zero energy modes at the two ends of the
system. Thus we realize the topological flatband in one dimension.

\begin{figure}[tbp]
\includegraphics[width=8cm]{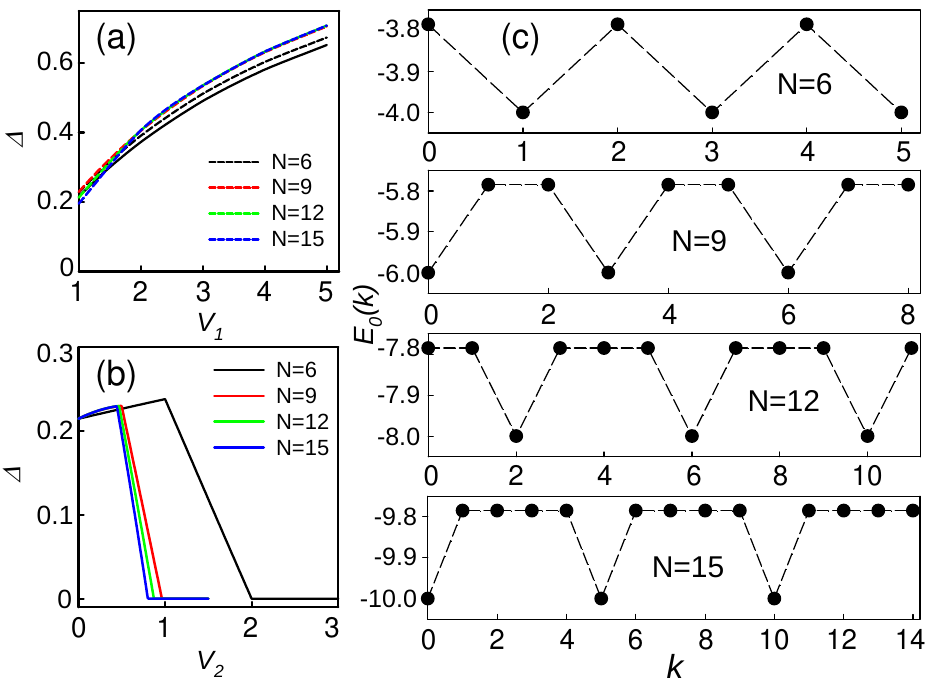}
\caption{(Color online) (a) The gap between the ground-states and the
excited states versus $V_1$ at $V_2=0$ (solid line) and $V_2=0.5$ (dashed
line). (b) the gap versus $V_2$ at $V_1=1$. (c) The ground-state energy of
each momentum sector at $V_1=1$ and $V_2=0$ on systems with different sizes.}
\label{fig2}
\end{figure}

\begin{figure}[tbp]
\includegraphics[width=8cm]{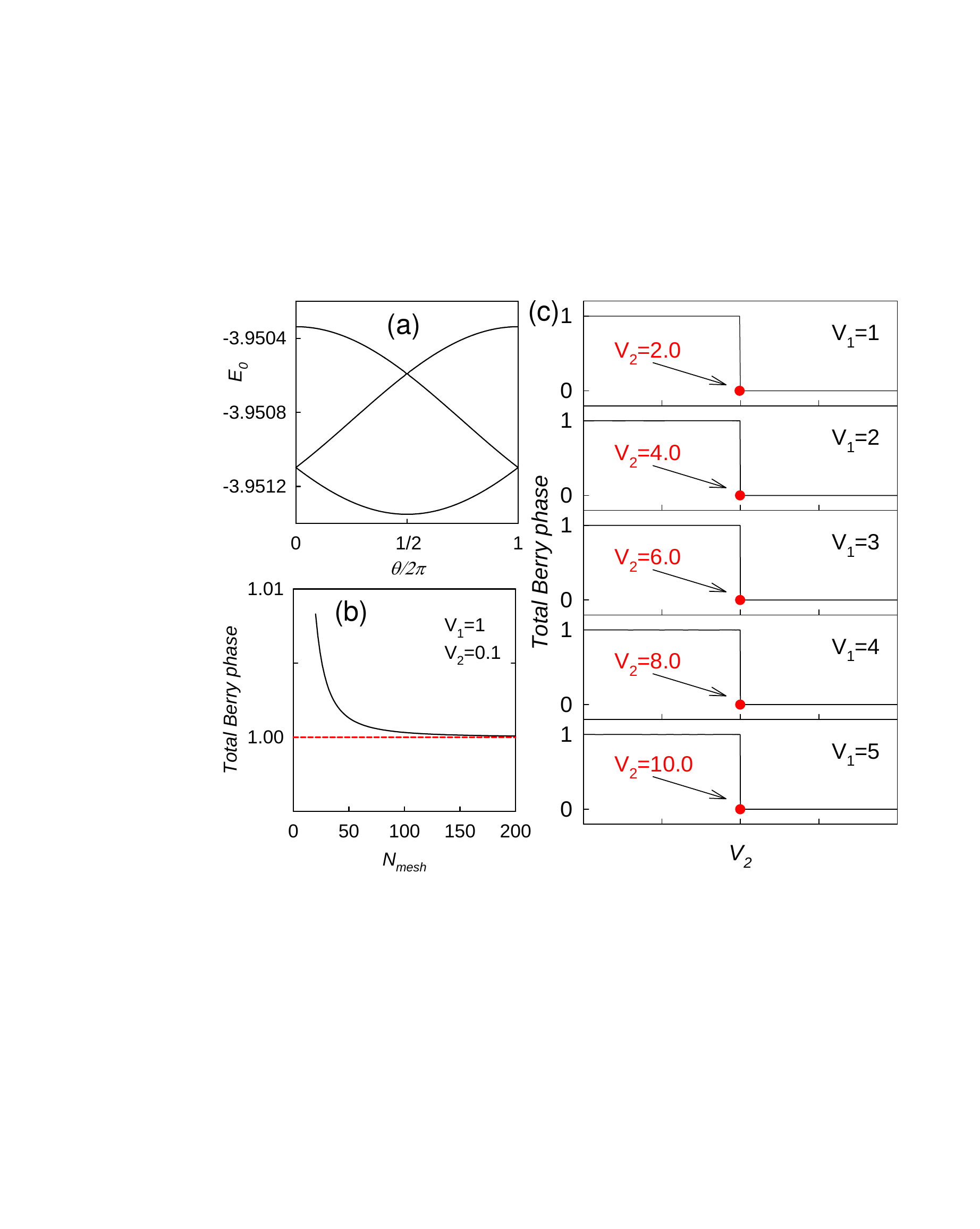}
\caption{(Color online) (a) The energies of the ground-states versus $%
\protect\theta$ at $V_1=1$ and $V_2=0$. (b) The total Berry phase versus the
number of divided meshes $N_{mesh}$. (c) The total Berry phase versus $V_2$
at different fixed $V_1$. Here the parameters are $A=B=1$ and $M=-1.99$ when
the band slightly departs the exact flatness.}
\label{fig3}
\end{figure}

\section{The effect of interaction}
 Next we study the effect of
interactions in the topological flatband. We firstly add
nearest-neighbouring (NN) and next-nearest-neighbouring (NNN) interactions
to Eq.(\ref{eq1}), which are written as,

\[
H_{I}=V_{1}\sum_{\langle i,j\rangle }n_{i}n_{j}+V_{2}\sum_{\langle \langle
i,j\rangle \rangle }n_{i}n_{j}
\]%
where $n_{i}=c_{i}^{\dagger }c_{i}+d_{i}^{\dagger }d_{i}$ is the total
number of electrons on site $\mathbf{r}_{i}$ and $V_{1}$, $V_{2}$ are the
strength of the interactions. We perform the exact diagonalization study of
the total Hamiltonian $H_{0}+H_{I}$ on a finite chain of $N$ sites with
periodic boundary condition. We denote the number of particles as $N_{p}$
and the filling factor of the topological flatband is $\nu =N_{p}/N$. We
have carried out the calculations at $\nu =1/3$, and identified the FTP in
which the ground-states are three-fold degenerate. We firstly glance at the
phase diagram in the $(V_{1},V_{2})$ plane, which is shown is Fig.\ref{fig1}
(c). By turning on $V_{1}$, the ground-state is three-fold degenerate and
the FTP emerges. The ground-state is separated from higher eigenstates by a
finite gap $\Delta $. As shown in Fig.\ref{fig2} (a), the value of the gap
increases with the strength of $V_{1}$, indicating that the FTP is more
stable at larger $V_{1}$. After turning on $V_{2}$, the value of the gap
vanishes and the FTP is destroyed at a critical value $V_{2c}$ [see Fig.\ref%
{fig2} (b)], which marks the boundary of the FTP in the phase diagram. We
determine the boundaries of the FTP on different sizes of the chain and find
that the region of the FTP is shrunk as the size increases. However
according to the results on the sizes we can access, it is reasonable to
deduce that the FTP exists in the thermodynamic limit.

The system under consideration has translational symmetry and the momentum
of the eigenstate is a good quantum number. Thus the Hamiltonian can be
diagonalized in each sector with the momentum $q=2\pi k/N$ ($k=0,1,...,N-1$%
), which allows us to examine the character of the low-energy spectrum in
the momentum space. In Fig.\ref{fig2} (c) we show the ground-state energy of
each momentum sector at $V_{1}=1$ and $V_{2}=0$ on systems with different
sizes. It has been shown that for these parameters the ground-states are
three-fold degenerate. Here we further demonstrate that the three states are
in different momentum sectors. If $k_{1}$ is the momentum sector for one of
the ground-state manifold, the other states should be obtained in the
sectors $k_{1}+N_{p}$ and $k_{1}+2N_{p}$ (module $N$). For the cases with
nonzero $V_{2}$ the results are similar. The gap between the ground-states
and the excited-states is clearly demonstrated in Fig.\ref{fig2} (c). And we
find that the value of the gap is independent of the size of the system for $%
V_{2}=0$ [the solid line in Fig.\ref{fig2} (a)], while for $V_{2}\neq 0$ it
is dependent [the dashed line in Fig.\ref{fig2} (a)].

To further confirm the existence of the FTP, we study the topological
property of the ground-states. It can be understood in terms of the total
Berry phase of the ground-states. By using the twisted boundary conditions,
it is expressed as \cite{berry1,berry2}
\[
\gamma =\sum_{j=1}^{3}\oint i\langle \psi _{\theta }^{j}|\frac{d}{d\theta }%
|\psi _{\theta }^{j}\rangle ,
\]%
where $\theta $ is the twisted boundary phase which takes values from $0$ to
$2\pi $ and $\psi _{\theta }^{j}$ are the corresponding many-body wave
functions of the three-fold degenerate ground-states. Since the flatband has
nontrivial topology, the total Berry phase of the ground-states is expected
to be $mod(\gamma ,2\pi )=\pi $. Each ground state shares the Berry phase $%
\pi /3$ averagely. When the bands are exactly flat, the energy of the
ground-states remains the same as $\theta $ varies. So to perform the
calculations, we adjust the parameter $M$ a little and make the band nearly
flat. As shown in Fig.\ref{fig3} (a), the ground-state energies are slightly
split and vary with $\theta $. Then we calculate the total Berry phase at
different $V_{1}$ on a chain with $N=6$ and show the result in Fig.\ref{fig3}
(c). It shows that the total Berry phase gets nontrivial value $\pi $ for
small $V_{2}$ and jumps to zero as $V_{2}$ is further increased. The
obtained critical values $V_{2c}$ are in good consistence with those from
the energy spectra. Moreover for other lattice sizes, the results are the
same. In the calculations, we divide the range of the boundary phase $%
[0,2\pi ]$ into $N_{mesh}=100$ meshes, which allows the results in an
acceptable precision. With the above methods, we also identify the FTP at $%
\nu =1/4$.

\begin{figure}[tbp]
\includegraphics[width=6cm]{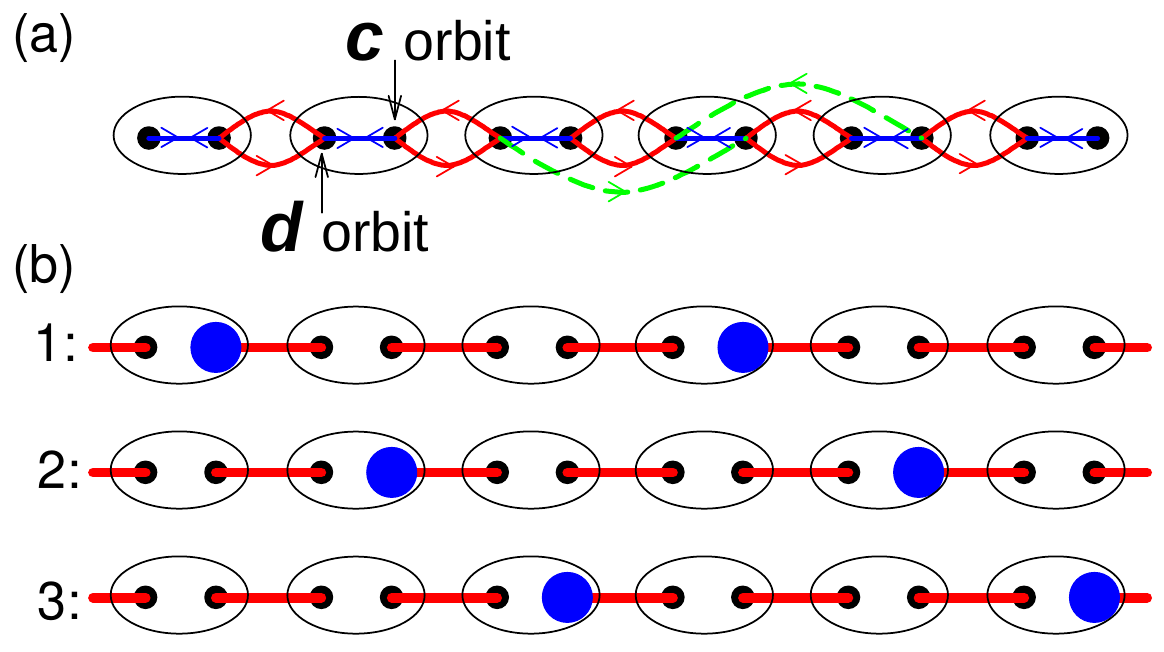}
\caption{(Color online) (a) The schematic representation of the resulting
Hamitonian after rotating the Pauli matrices. The hopping amplitudes of the
blue, red and green bonds are $M+2B$, $-(B+A)$ and $-(B-A)$ respectively.
(b) The three degenerate ground-states at $\protect\nu=1/3$.}
\label{fig4}
\end{figure}

\section{Mapping to the Su-Schrieffer-Heeger model}
The Hamiltonian
described by Eq.(\ref{eq1}) can be obtained through a dimensional reduction
from the Bernevig-Hughes-Zhang model which describes the two-dimensional
topological insulator HgTe \cite{ssh1}. Performing a cyclic permutation of
the Pauli matrices $\sigma _{z}\rightarrow \sigma _{x}$, $\sigma
_{x}\rightarrow \sigma _{y}$, and $\sigma _{y}\rightarrow \sigma _{z}$ in
Eq. (\ref{eq1}), the Hamiltonian becomes,
\begin{eqnarray}
H_{0}^{\prime } &=&\sum_{i}(M+2B)(c_{i}^{\dagger }d_{i}+d_{i}^{\dagger
}c_{i}))  \label{eq2} \\
&-&\sum_{i,\hat{x}}([B+sgn(\hat{x})A]c_{i}^{\dagger }d_{i+\hat{x}}+[B-sgn(%
\hat{x})A]d_{i}^{\dagger }c_{i+\hat{x}}),  \nonumber
\end{eqnarray}%
where the hopping amplitudes are all real. If the two orbits on each site
are viewed as two separate sites, it describes the free electrons on a chain
with NN and next-next-nearest-neighbouring hoppings, as shown in Fig.\ref{fig4}(a). In the flatband
case of $A=B=-M/2$, it can be simplified as,
\[
H_{0,flat}^{\prime }=-2A(\sum_{i}c_{i}^{\dagger }d_{i+1}+d_{i+1}^{\dagger
}c_{i}),
\]%
which is the Su-Schrieffer-Heeger model in the limit case where only NN
hopping exists and every other bond is broken completely \cite{ssh2,ssh3}.
The model is solvable by setting $\gamma _{i,\pm }=(c_{i}\pm d_{i+1})/\sqrt{2%
}$. Then $H_{0,flat}^{\prime }=-2A(\sum_{i}\gamma _{i,+}^{\dagger }\gamma
_{i,+}-\gamma _{i,-}^{\dagger }\gamma _{i,-})$ and then we have the two
flatband spectra as shown in Fig.\ref{fig1}(a). At $\nu =1/3$, the lowest
energy state is highly degenerate. In the presence of the NN interaction in
Eq. (\ref{eq1}), there are three configurations of the ground states where
the NN interaction between the electrons can be minimized, $\left\vert
g_{,\alpha }\right\rangle =\Pi _{n}\gamma _{3n+\alpha ,+}^{\dag }\left\vert
0\right\rangle $ with $\alpha =0,1,2$. These three states $\left\vert
g_{,\alpha }\right\rangle $ are the charge-density-wave as depicted in Fig. %
\ref{fig4}(b), which breaks the translational symmetry. However, we found
that the lowest energy states in each momentum sector $K$ has a uniform
density distribution of electron, which is \ in consistence with the
translational symmetry in Eq.(\ref{eq1}). The states are the linear
combination of the three degenerate states $\left\vert g_{,\alpha
}\right\rangle $. Utilizing the lattice translational operator, we can
construct the three states as eigenstates of the momentum \ $K:\left\vert
g,K\right\rangle =\frac{1}{\sqrt{3}}\sum_{\alpha }e^{i\alpha K}\left\vert
g_{,\alpha }\right\rangle $ where $K=2n\pi /3$ ($n=0,1,2$). Thus these three
states are those in the presence of interactions. It can be checked as a
limit of the nearly flatband, in which the density distribution of electron
is always uniform. Therefore the ground-states show three-fold degeneracy.
After the NNN interaction is turned on, though the electrons begin to
interact with each other, the three configurations still have the same
lowest energies for small $V_{2}$ and the three-fold degenerate
ground-states persist for $V_{2}<V_{2c}$. In particular with the above
picture $n-$ fold ($n$ other than three or four) degenerate ground-states
can be generated by introducing proper interactions.

\begin{figure}[tbp]
\includegraphics[width=8cm]{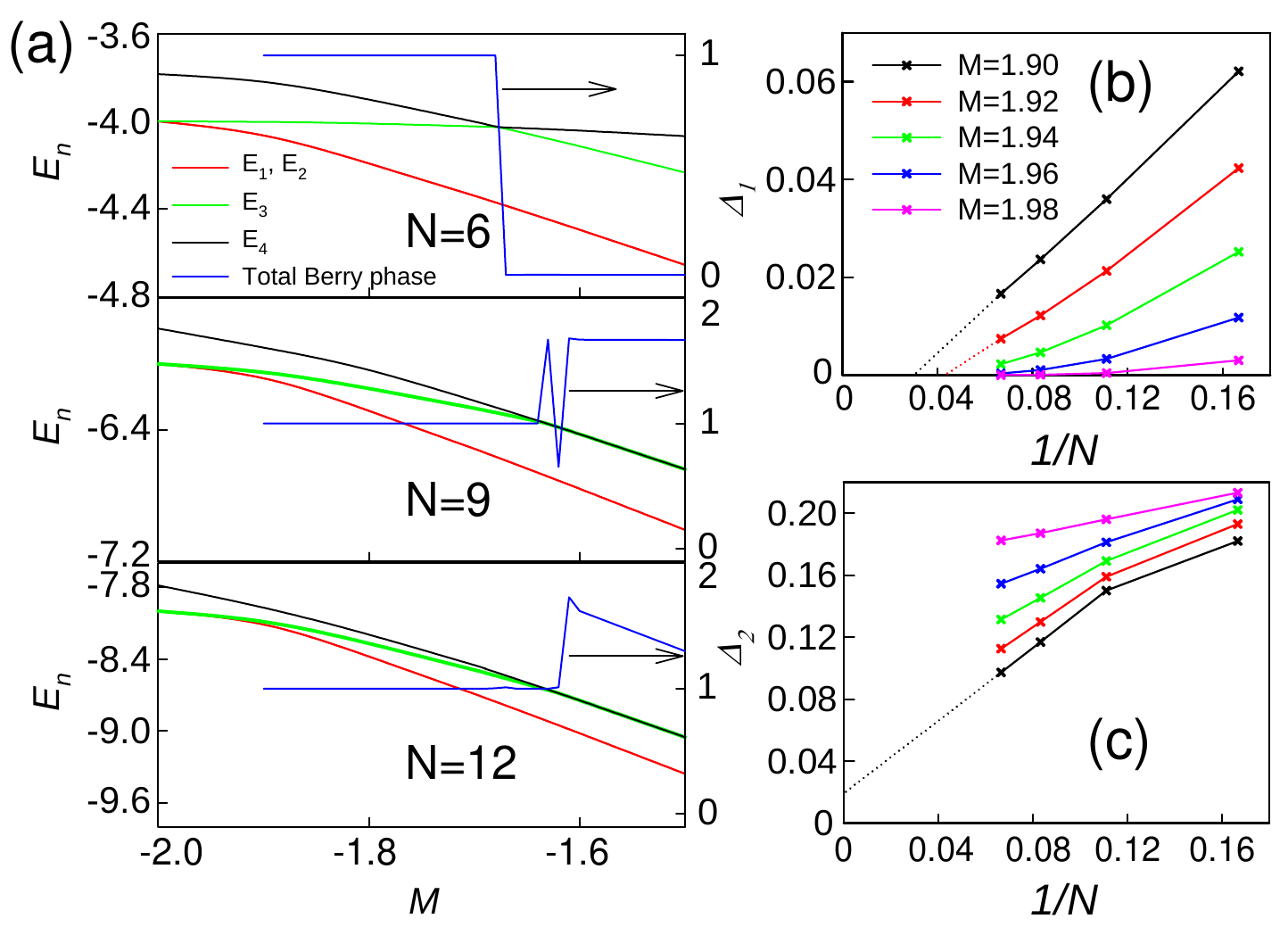}
\caption{(Color online) (a) The energies of the lowest four states and the
total Berry phase of the lowest three states versus $M$. (b) and (c) the
finite-size analysis of $\Delta_1$ and $\Delta_2$ respectively. Here $%
\protect\nu=1/3$.}
\label{fig5}
\end{figure}

\section{The FTP in the nearly flatband}
Up to now we have established the
existence of the FTP in topological flatband. It is natural to ask whether
the phase persists in the nearly flatband. In the following we let the band
dispersing by tuning the parameter $M$ and study the properties of the
ground-states. As shown in Fig.\ref{fig5} (a), when $\delta M=2-M\neq 0$ the
the energy of three lowest energy states are split. An energy gap $\Delta
_{1}$ appears between the lower two-fold degenerate states and the upper
one, whose value increases with $\delta M$. Meanwhile the energy gap $\Delta
_{2}$ separating the three lowest energy states from the fourth lowest
energy state decreases with $\delta M$ and vanishes at a critical value $%
\delta M_{c}$. Moreover beyond $\delta M_{c}$ the total Berry phase of the
three states with the lowest energies is no longer quantized to $\pi $,
which implies the occurrence of a topological quantum phase transition. Now
the question becomes whether the three lowest states are still degenerate
for $0<\delta M<\delta M_{c}$ in the thermodynamic limit. To this end, the
finite-size analysis of two energy gaps $\Delta _{1}$ and $\Delta _{2}$ at
several small $\delta M$ is presented in Figs. \ref{fig5}(b) and (c). The
results demonstrates that $\Delta _{1}$ tends to vanish while $\Delta _{2}$
survives in the thermodynamic limit, i.e., for a large $N$ limit. So the
splitting of the degeneracy of the lowest energy states may be due to the
finite-size effect and the FTP persists in the nearly flatband.

\section{Conclusions}
We have studied the FTP in a 1D interacting
topological flatband model using exact diagonalization. By introducing the
NN and NNN interactions, we identify the FTP at $\nu =1/3$, in which the
ground-states are three-fold degenerate. Moreover the degenerate
ground-states are in different momentum sectors and are equally spaced with
the interval of $N_{p}$. With the above method, we have obtained the phase
diagram at $\nu =1/3$ in the $(V_{1},V_{2})$ plane. We also study the total
Berry phase of the low-energy states and find that it gets a quantized value
$\pi $ when the system is in the FTP. The phase boundary from the total
Berry phase is in good consistence with that from the low-energy spectrum.

Existence of the 1D FTP is closely related to the underlying physics in
dimerized polyacetylene \cite{ssh1,ssh2,ssh3}. One explicit consequence of
the 1D FTP is the charge fractionalization of the quasi-particles or
excitations. Following Kivelson and Schrieffer \cite{Kivelson-82}, in the
3-fold degenerate ground-states, the domain walls or solitons possess the
fractional charges of $e/3$, which can be changed by an integer by adding
electrons or holes. In our case, suppose a domain wall is formed between two
of the three degenerate ground-states. The domain wall can move freely on
the chain. When we move the particle one unit, the domain wall moves three
units. Since the change in electric dipole moment can be calculated in two
equivalent ways, i.e., through the particle motion $e(+1)$ and through the
domain wall motion $Q(+3)$, we have $Q=e/3$, which is the fractional
excitation associated with the domain wall of the 1D FTP. For strongly
interacting systems, it can be also understood very well from the
bosonization and the theory of macroscopic polarization by calculating the
Berry phases \cite{Aligia-05prb}.

Now we can have a systematic understanding of the FTPs on lattices without
net magnetic field in all three dimensions. For the 1D case, the flatband
has a nontrivial Berry phase and the excitations with fractional charges
characterize the FTP. For the 2D case, the FTP is found in interacting
electrons of flatband with a nontrivial Chern number. Its feature is an
almost multi-fold degenerate incompressible ground-states with fractional
Hall conductance, which is similar to FQHE. Also by combining the two
decoupled FTPs formed by spin up and down electrons, we have fractional
quantum spin Hall effect or fractional TI with time-reversal symmetry. For
the 3D case in nearly flatband characterized by a nontrivial $Z_{2}$
topological index and in the presence of repulsive interactions, a 3D
fractional TI can be generated. Thus FTP can be realized in interacting
electrons with topologically non-trivial bands. The common features are the
multi-fold degeneracy of the ground states and topological invariants of all
degenerated ground states.

Due to the rapid development of the field
of cold atoms \cite{cold1}, it is very hopeful to realize the 1D FTP on
state-dependent optical lattice with laser-assisted tunneling between adjacent sites \cite{cold2,cold3,cold4}. To simulate our Hamiltonian, atoms with four internal states $c_{1,2}$ and $d_{1,2}$ are required, such as $^{6}$Li et al.. Then in the state-dependent optical lattice, atoms in states $c_{1}$ and $d_{1}$ are at positions ${\bf r_{1}}=2n$ and atoms in states $c_{2}$ and $d_{2}$ are at positions ${\bf r_{2}}=2n+1$, where the lattice constant is $\lambda/4$ with $\lambda$ the wavelength of the laser generating the optical lattice. We can choose the optical potential large enough to prohibit direct tunneling between neighboring sites. The hoppings in the model are induced by additional lasers driving Raman transitions between different states. By setting different resonance frequencies for different kind of hopping, the non-interacting Hamiltonian can be realized using proper additional lasers. Since the separation between neighboring sites is half the original lattice constant, significant NN interactions can be generated. Also due to the fact that on-site interactions don't affect the FTP, it is very possible to realize the FTP in this scheme. Finally it would be exciting to find real
materials exhibiting the FTP.

\section{Acknowledgements}
The authors would like to thank Z.-C. Gu, Hua Jiang, D. N.
Sheng, Jun-Ren Shi and Yi-Fei Wang for helpful discussions. HG is supported
by FOK YING TUNG EDUCATION FOUNDATION and NSFC under Grant No. 11104189; SS
is supported by the Research Grant Council of Hong Kong under Grant No.
N\_HKU748/10; SF is supported by the Ministry of Science and Technology of
China under Grant Nos. 2011CB921700 and 2012CB821403, and NSFC under Grant
No. 11074023.

\end{document}